# Predicting a Business' Star in Yelp from Its Reviews' Text Alone

Mingming Fan

College of Computing and Informatics

University of North Carolina, Charlotte

Charlotte, USA

mfan@uncc.edu

Maryam Khademi

Donald Bren School of Information and Computer Sciences

University of California, Irvine

Irvine, USA

mkhademi@ics.uci.edu

*Abstract*—Yelp online reviews are invaluable source of information for users to choose where to visit or what to eat among numerous available options. But due to overwhelming number of reviews, it is almost impossible for users to go through all reviews and find the information they are looking for. To provide a business' overview, one solution is to give the business a 1-5 star(s). This rating can be subjective and biased toward users' personality. In this paper, we predict a business' rating based on user-generated reviews' texts alone. This not only provides an overview of plentiful long review texts but also cancels out subjectivity. Selecting the restaurant category from Yelp Dataset Challenge [1], we use a combination of three feature generation methods as well as four machine learning models to find the best prediction result. Our approach is to create bag of words from the top frequent words in all raw text reviews, or top frequent words/adjectives from results of Part-of-Speech analysis. Our results show Root Mean Square Error (RMSE) of 0.6 for the combination of Linear Regression with either of the top frequent words from raw data or top frequent adjectives after Part-of-Speech (POS).

## I. INTRODUCTION

More than ever before, people's decisions of where to visit or what to eat are subject to other people's opinions. The internet has become the ultimate trove of the opinions of many, many people. Today, websites like Yelp have turned to a vast database for places and restaurants that include reviews and opinions written by everyday people. This crowdsourcing method of extracting satisfaction has succeeded in providing different opinions about a certain service.

User-generated reviews are usually inconsistent in terms of length, content, writing style and usefulness because they are written by unprofessional writers. Important information can be easily obscured unless users are willing to spend a great deal of time and effort on reading the reviews thoroughly. A common solution to provide a brief overview is to show overall rate of a business in form of 1-5 star(s). While "Yelp ratings are often considered as a reputation metric for businesses" [2], they may suffer from subjectivity and being biased toward users' personality.

Fig. 1 demonstrates bias in two reviews selected from Yelp website. As can be seen, there are two users (Michelle and Clif) who wrote about "Providence", a restaurant in LA area. Both users seem to be very pleased with their experience because they described it with multiple positive words such as "perfection", "must go", "great treat", "tasted great", etc. However, the first user gave five stars to the restaurant whereas the second user gave three stars. This became our motivation to predict a business' rating based on its reviews text alone to reduce the bias of the reviewers.

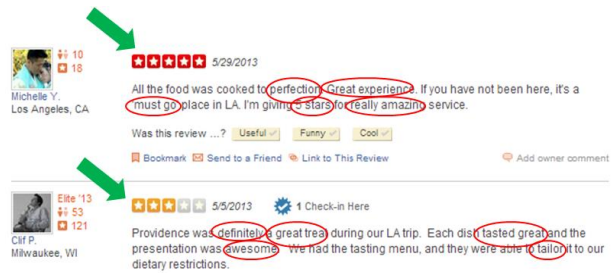

Fig. 2. An example of bias in Yelp website's reviews

## II. RELATED WORK

A lot of efforts have been invested in information extraction from web [3], opinion mining [4], and in particular review mining [4]–[6]. Dave et al [6] developed a system to find review's tags and associated sentiment score with them. Lee et al. [7] introduced a system that users were able to add tags with a negative/positive sentiment to a review. Turney [8] and Pang et al. [9] analyzed review texts using machine learning algorithms and n-gram techniques to determine the sentiment orientation of the phrases. Our work relates to this group of literature in the sense that it is opinion mining but we did not use sentiment analysis to tag positivity/negativity of the reviews.

Yatani et al. [10] and Huang et al. [11] designed different interfaces for Yelp that show top frequent adjectives used to describe a business. These interfaces also visualized overall sentimental scores of reviews in various colors. The authors did not provide any evidence to show whether using adjectives are more effective than other words. It is also not clear whether using sentimental scores have more advantages over raw reviews' text. In contrast, we focused on generating different features and analyzing how these features affect the performance of our prediction models to address the mentioned questions.

To our knowledge, there has been little research on Yelp's data. One example of that is Potamias [2] who studied how

Yelp ratings evolve over time. His main finding was that on average the first ratings that businesses receive overestimate their eventual reputation. He showed that the first review that a business receives averages 4.1 stars, while the 20th review averages just 3.69 stars. This 'warm-start bias', which may be attributed to the limited exposure of a business in its first steps, may mask analysis performed on ratings and reputational ramifications. Through this work, we hope to encourage more researchers to study and analyze the Yelp dataset.

### III. METHODOLOGY

#### A. Data

We collected our data from Yelp Dataset Challenge. This dataset contains 11,537 businesses, 8282 check-in sets, 43873 users, and 229907 reviews. We target "Restaurant" as the most famous category amongst others. There are 4243 restaurants in total. We randomly chose 1000 of them and analyzed 35645 text reviews written about them. To evaluate our results, we divided our data into 90% training and 10% testing datasets using 10-fold cross validation. For training dataset, we used both text review and business' star. For testing, we used our models to predict the business rating and then compared it with the actual rating that we had to evaluate the accuracy of our model.

As part of our exploratory analysis, we found that words such as "food" (17%), "good" (17%), and "place" (16%) have the highest frequency amongst all. Fig. 2 illustrates the 12 most repeated words in all reviews of our selected 1000 restaurants.

#### B. Feature Selection

To form the feature vector, first we formed bag of words from text reviews of all restaurants. Second, we picked the top K frequent words. Finally, we calculated the frequency of each top K words in all reviews of each business. Table 1 shows an example of how our feature matrix looks like.

We generated three feature vectors including the features extracted from raw data as well as those that were engineered. Our three feature generation methods include: (i) baseline; (ii) feature engineering I, and (iii) feature engineering II.

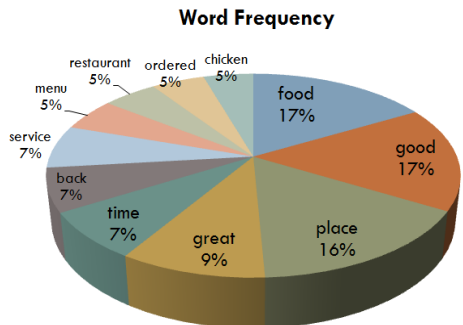

Fig. 2. Top 12 frequent words used to describe our 1000 selected restaurants

TABLE 1: THE FEATURE MATRIX

|  | Word #1 | . . | Word #K | Star |
|---|---|---|---|---|
| **Business #1** | Frequency of word #1 |  | Frequency of word #K | 4 |
| . . |  |  |  | . . |
| **Business #N** |  |  |  | 3 |

*1) Baseline*

We analyzed the raw data to choose the top K frequent words used in all reviews. This array of words became our feature vector. Then we went through the reviews of each 1000 restaurants and counted the number of times that each member of our feature vector was used to describe that restaurant. Finally, we divided the number of occurrence with the total number of occurrence of all the top K words to calculate the frequency of our feature vector members using this formula:

$$\text{freq}(i) = \frac{x_i}{\sum_{i=1}^{K} x_i} \quad (1)$$

where $x_i$ is the number of times that i-th top K frequent word appeared in the review of a restaurant. We used K equal to 30, 50, 100, 200, 300, 500, and 1000 to examine which one gives the least error. Given that we analyze N restaurant(s) in total, with a feature vector of size K, the training matrix is of size N x K.

*2) Feature Engineering II*

Our first hypothesis was that Part-of-Speech (POS) might help figure out the most representative words as our feature vector. "Part-of-Speech can effectively analyze syntactic/semantic structures of English sentences and provides phrase structures and predicate-argument structure" [12]. We did Part-of-Speech analysis per sentence. Using the results of Part-of-Speech analysis, we selected the top K frequent words amongst all. We picked different number of the top K frequent words including: 30, 50, 100, 200, 300, 500, and 1000.

*3) Feature Engineering II*

Our second hypothesis was that creating feature vector from the top K frequent adjectives might yield better result compared to the top K frequent words after POS. The reason is that adjectives are the most commonly used type of words to describe positivity or negativity. After finishing POS, we extracted the adjectives only and picked the top K frequent adjectives for the same K values used in previous methods, i.e., 30, 50, 100, 200, 300, 500, and 1000.

#### C. Learning Methods

Having the bag of words, we treat the problem of predicting a business star as a regression problem. We choose four learning models: (i) Linear Regression; (ii) Support Vector Regression; (iii) Support Vector Regression with normalized features; and (iv) Decision Tree Regression.

### IV. EVALUATION AND RESULTS

Because our goal is to predict the business' star using regression model, we use the Root Mean Square Error to

quantify our error, instead of using accuracy. The equation is shown below:

$$RMSE = \sqrt{\frac{1}{n} * \sum_{j=1}^{n}(y_j - \hat{y_j})^2} \quad (2)$$

Recall we had three feature generations methods: (i) Baseline: Top frequent words directly from the raw text reviews; (ii) Feature Engineering I: Top frequent words after doing Part-of-Speech analysis on all reviews; and (iii) Feature Engineering II: Top frequent adjectives after doing Part-of-Speech analysis on all reviews. We also had four learning models: Linear Regression, Support Vector Regression (with and without normalized features), and Decision Tree Regression. Specifically for SVR, we tried both normalized/not normalized features to see whether normalization has an effect on the results. Fig. 3-5 show the results of our five learning models on three feature generation methods.

From Fig. 3-5, we conclude the following findings:

*1)* No matter which feature generation methods we use, the Linear Regression model performs the best. This implies that the relationship between the features and target in our data might be linearly correlated.

*2)* As the number of features increases, the RMSE drops first and then jumps up. The best performance for different models appears while having around 50 to 100 features.

*3)* Decision Tree's performance is robust with respect to the number of features. The performance is very close to that of Linear Regression in most instances. In our test, we also noticed that Decision Tree is super-fast.

*4)* Normalization helps improve the performance of Support Vector Regression.

We also compared the best performance of the four different models under our three different feature generation methods (See Fig. 6 and Table 2). As can be seen, using the features directly from the raw data has almost equivalent power as fine engineered features. Also, the Linear Regression learning model performs the best for all feature generation methods.

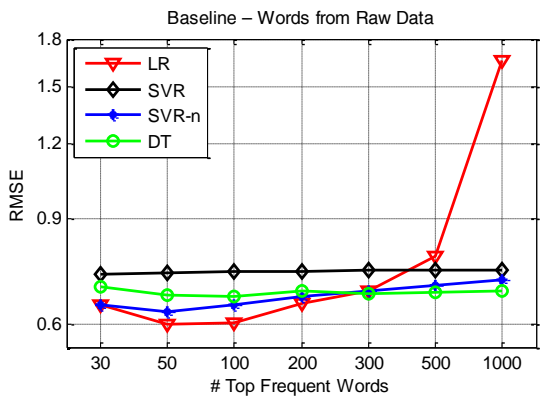
Fig 3. Results of learning methods for baseline

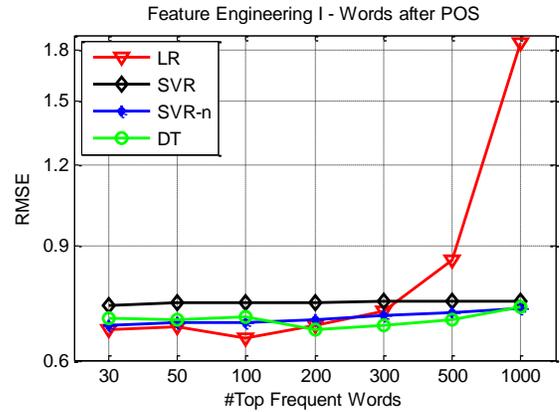
Fig 4. Results of learning methods for feature engineering I

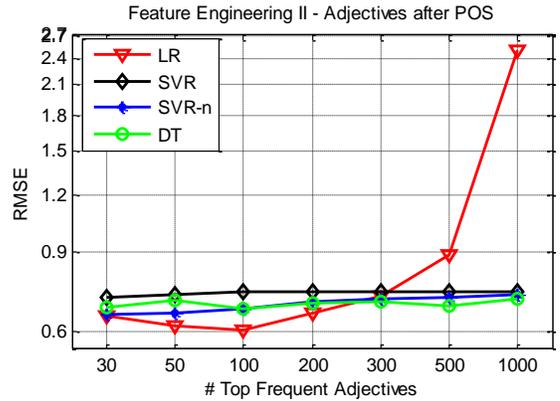
Fig 5. Results of learning methods for feature engineering II

TABLE 2: BEST RESULTS OF THE FOUR MODELS WITH OUR THREE FEATURE GENERATING METHODS; LEAST RMSE'S ARE HIGHLIGHTED.

| Feature Selection Method | Learning Model | RMSE |
|---|---|---|
| **Baseline (Top Frequent Words from Raw Data)** | Linear Regression | 0.6014 |
| | Support Vector Regression | 0.7278 |
| | Support Vector Regression-n | 0.6296 |
| | Decision Tree | 0.6689 |
| **Feature Engineering I (Top Frequent Words after POS)** | Linear Regression | 0.6488 |
| | Support Vector Regression | 0.7298 |
| | Support Vector Regression-n | 0.6791 |
| | Decision Tree | 0.6697 |
| **Feature Engineering II (Top Frequent Adjectives after POS)** | Linear Regression | 0.6052 |
| | Support Vector Regression | 0.7135 |
| | Support Vector Regression-n | 0.6766 |
| | Decision Tree | 0.6766 |

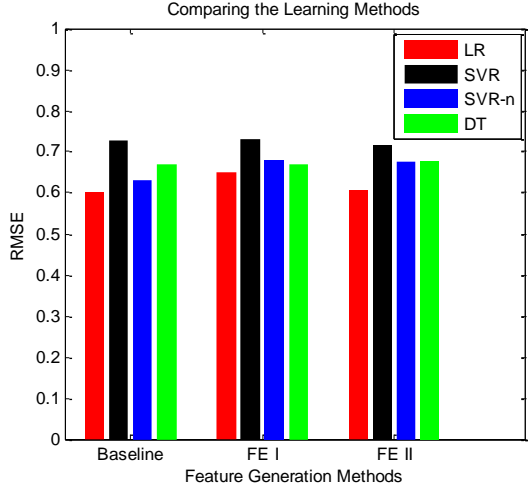

Fig 6. comparing the four learning models for our three feature generation methods: baseline, feature engineering I, and feature engineering II

## V. SCALABILITY

We analyzed the time complexity of the feature generation methods. Here are the annotations:

- $N_{all_{words}}$: # number of all the words in all reviews of restaurants.

- $N_{all_{adjs}}$: # number of all the adjectives in all reviews of restaurants.

- $N_{biz}$: # number of businesses in all JSON files (Yelp dataset)

- $N_{rev}$: # number of reviews in all JSON files (Yelp dataset)

- $N_{top_{adjs}}$: # top frequent appeared adjectives in all reviews

The time complexity of each part is analyzed as below:

1. Building HashMap of business' id and star: $O(N_{biz} \times \log N_{biz})$
2. Building HashMap of business' id and all its reviews: $O(N_{rev} \times \log N_{rev})$
3. Selecting top adjectives: $O(N_{all_{words}} \times \log N_{all_{adjs}})$
4. For each business, counting all top adjectives' frequency: $O(N_{all_{words}} \times \log N_{top_{adjes}})$

Because $N_{all_{words}} > N_{all_{adjs}} > N_{top_{adjs}}$ and $N_{rev} > N_{biz}$, the final time complexity is $O(N_{all_{words}} \times \log N_{all_{adjs}} + N_{rev} \times \log N_{rev})$. Using HashMap while processing large amount of text reviews accelerates making the bag of words and search for all words' frequencies because the search time complexity is $O(\log N)$.

About space complexity, we didn't process all businesses' data at once. Instead, we only focused on analyzing a specific business' reviews at a time. We therefore reduced the amount of data to be loaded in memory and needed to be processed at each step. In instance a business had huge amount of text reviews (e.g. more than 35000 reviews), we separated its reviews into smaller chunks (i.e., at each time, we only processed one part of it and wrote the intermediate results to the disk). After all processing were done, we merged the intermediate parts together.

## VI. DISCUSSION

In this paper, we tried different feature generation methods (i.e., word frequency directly from the raw text review, word frequency after Part-of-Speech, and adjective frequency after Part-of-Speech) as well as four learning models (i.e., Linear Regression, Support Vector Regression (with and without normalized features) and Decision Tree Regression). We found that:

For different feature generation methods, there were no significant differences in RMSE. The raw data had equivalent power as our feature engineering methods. Although calculating word frequency directly from raw data might have the issue of root form (e.g. treat "like", "likes", "liked" as three features) and therefore some features may be potentially correlated with each other, this approach never abandoned any important information. In contrast, calculating word or adjective frequency after POS might remove informative features out. That might be an explanation to why extracting the features directly from raw data was very competitive to the other two engineered feature methods.

For learning models, Linear Regression performs the best in general. Because Linear Regression assumes the linear relationship between features and target, we can infer that for our data the features and target might be linearly correlated.

Too many features poison Linear Regression. In Fig. 3-5, we can see that once number of features exceeds 500, the performance of Linear Regression drops. The potential reason is that Linear Regression is sensitive to outliers. The more features we include, the higher the chance of including more outliers.

The performance of Decision Tree Regression is robust with respect to the number of features. And the overall performance is competitive compared with the other models, though it is not the best. More importantly, Decision Tree Regression training is super fast. Therefore, it would be a good compromise if there is a time limitation.

Normalization of features is helpful for Support Vector Regression model.

## VII. CONCLUSION

Given the increasing amount of peer-reviewed texts showing up in social network and recommender systems, the diversity and quality of text reviews will make important information drown in the noise. In this paper, we focused on Yelp Dataset Challenge with enormous amount of customers' reviews to predict a business's star only from its customers' text reviews. Our motivation was to remove the bias of stars given by different users.

We tried three feature generation methods and applied four machine learning models to them including Linear Regression, Support Vector Regression (with and without normalized features), and Decision Tree Regression. The final result showed that Linear Regression with top frequent adjective after POS as well as Linear Regression with the top frequent words extracted from raw data performed the best among other combinations.

As future work, we plan to generalize our model by considering other business categories (e.g. fashion, beauty, etc) to see whether it affects our model's accuracy.